\documentclass[aps,prb,twocolumn,floats,showpacs,superscriptaddress,nofootinbib]{revtex4-2}
\usepackage{graphicx,epsfig}
\usepackage{amssymb,amsmath,amsfonts}
\usepackage[title,titletoc,toc]{appendix}
\usepackage[pagebackref=false,colorlinks,linkcolor=magenta,citecolor=blue,urlcolor=magenta]{hyperref}
\usepackage{braket}
\usepackage{physics}
\usepackage{bm}
\usepackage[utf8]{inputenc}

\begin{document}

\title{Non-perturbative theory of valley splitting in Si qubits from variational wave function: periodic effects of shear strain and asymptotic freedom in the wiggle-well potential}

\author{J. L. P. Steinschuld}
\affiliation{2nd Institute of Physics C, RWTH Aachen University, 52074 Aachen, Germany}
\author{H. J. Bluhm}
\affiliation{2nd Institute of Physics C, RWTH Aachen University, 52074 Aachen, Germany}
\author{S. A. Jafari}
\affiliation{2nd Institute of Physics C, RWTH Aachen University, 52074 Aachen, Germany}
\email{akbar.jafari@rwth-aachen.de}
\date{\today}

\begin{abstract}
Valley splitting sets the energy scale at which spin and valley degrees of freedom hybridize in silicon quantum-well qubits, 
but its sensitivity to interface structure makes it difficult to predict. Using the valleyor basis, we formulate a two-band effective-mass model 
for intervalley coupling in a finite quantum well and develop an analytical framework for treating non-perturbative effects of 
shear strain and wiggle-well potentials. For a $z$-independent coupling $V_s\tau_2$, corresponding to uniform shear strain, 
we construct a variational state from the exact plane-wave valleyors and a hard-wall envelope. This yields closed-form expressions for the two 
lowest states and their splitting, including the characteristic oscillatory $|\sin(k_{\rm min}L)|$ dependence. The variational result 
agrees with exact diagonalization to within $\sim1\%$ for realistic couplings, and provides a simple prescription for tuning the 
shear strain away from the nodes to enhance the valley splitting. We further show that finite barrier heights primarily renormalize the 
effective width of the quantum well while preserving the oscillatory dependence. For a wiggle-well coupling $V_w\cos(k_wz)\tau_1$, the valleyor representation reveals a crossover near the $k_w=2k_1$ resonance from wiggle-dominated splitting to orbital quantization. 
Remarkably, at high wiggle-well amplitudes, the empty-box energy scale predominantly dictates the near-resonant valley splitting, while the wiggle-well potential enters only as a sub-leading correction. This behavior can be viewed as asymptotic freedom for spin qubits.
Away from resonance, the splitting exhibits a broadly peaked resonance  with diffraction-like side-lobes as a function of detuning.
\end{abstract}

\maketitle

\section{Introduction}
\label{sec:intro}

Valley splitting -- the energy gap that lifts the twofold valley degeneracy of
the conduction-band minima near the $X$ point of silicon -- is a key parameter
for silicon-based spin qubits~\cite{Burkard2023}, since it sets the energy scale below which
spin and valley states can hybridize and compromise qubit coherence
\cite{cai2023}. Because the splitting arises from
atomic-scale details of the confining interface, it is highly sensitive to
interface roughness, alloy disorder, and growth conditions, and can vary
substantially between nominally identical devices
\cite{marcks2025}. Predicting valley splitting has
motivated theoretical work ranging from atomistic tight-binding
\cite{mcjunkin2022} and density-functional calculations \cite{cvitkovich2026} to
effective-mass and $k\cdot p$ treatments that capture the essential valley
physics with a small number of parameters \cite{feng2022,Prentki2026}.

The microscopic origin and magnitude of $\Delta E$ are intimately connected to the atomic structure of the Si/SiGe interface. Early theoretical work established that intervalley coupling is generated by the rapidly varying potential at the heterostructure interface \cite{Friesen2007,Saraiva2009}, while experiments demonstrated that atomic steps and interface disorder can substantially modify the resulting valley splitting \cite{Goswami2007}. In particular, the valley splitting in Si/SiGe quantum dots can vary considerably between devices~\cite{Borselli2011}, reflecting the sensitivity of the intervalley matrix elements to atomic fluctuations~\cite{PaqueletWuetz2022,Prentki2026}, gate voltages~\cite{Yang2013} and electromagnetic fields at the interface~\cite{Lima2023,Hosseinkhani2020}. Atomic fluctuations give rise to random valley splitting, with a significant tail extending toward zero valley splitting, which can be detrimental to the shuttling of spin qubits~\cite{PaqueletWuetz2022}.

The ``wiggle-well'' (WW) \cite{mcjunkin2022} modulates the Ge concentration, or
equivalently the confining potential, periodically along the growth direction.
A uniform shear strain $\varepsilon_{xy}$ provides another symmetry-allowed
intervalley coupling \cite{jafari2026,Hensel1965}. We describe both mechanisms within a
two-band effective-mass model formulated in the ``valleyor'' basis
\cite{jafari2026}. The two components of a valleyor are associated with the two-fold degeneracy at the 
$X$ point and form an intrinsic two-dimensional valley degree of freedom, analogous in structure to
the two-component description used for spin, but associated with the extra
representation of the non-symmorphic space group at the $X$ point rather than
with physical spin. This basis therefore provides a natural framework in which
the valley degree of freedom is treated explicitly as a two-component internal
degree of freedom, with intervalley perturbations represented by Pauli matrices
in valley space. 

The purpose of this paper is to demonstrate that, in addition to providing a faithful representation of the silicon valleys, 
the valleyor basis offers significant computational advantages, allowing us to derive closed-form expressions that capture the non-perturbative effects of strain and detuned wiggle-well potentials. 
This non-perturbative treatment reveals a periodic dependence of the valley splitting on the shear strain for a fixed heterostructure width $L$, 
a behavior that cannot be captured by perturbative treatments of the strain effects. The practical implication of this result is that, 
for each heterostructure width $L$, the shear strain should be tuned to an antinode of this sinusoidal dependence to maximize the valley splitting.
 
Our valleyor basis and its analytical capabilities also provide insight into the role of the wiggle-well potential by identifying 
two distinct regimes, in which the wiggle wavelength determines whether the valley splitting is dominated by the box-quantization 
energy scale of the heterostructure or by the wiggle amplitude. 
In particular, we find that, quite surprisingly and counterintuitively, when the wiggle-well potential amplitude 
is very strong and its wavelength is tuned to the $2k_1$ resonance, or deviates only slightly from it, the valley splitting 
remains dominated by box quantization. The wiggle-well potential provides only a subleading correction in this regime, 
in agreement with the numerical results.

The valleyor formulation is particularly useful when the longitudinal momentum
$k$ is well defined. For a $z$-independent coupling, the Hamiltonian is diagonal
in momentum, so that the plane-wave problem can be solved exactly before the
finite-well boundary conditions are imposed variationally. We construct trial
states from the exact eigenvalleyors at $\pm k_{\rm min}$, dressed by a
hard-wall envelope. The resulting ansatz reproduces the numerical
$|\sin(k_{\rm min}L)|$ dependence of the valley splitting and approaches the
uncoupled-well limit as $V_s\to0$. We benchmark both the splitting and the
wave function against exact diagonalization, and then apply the same framework
to the WW problem. Finally, we analyze the effects of a confining potential with a
finite depth. 

\section{Model}
\label{sec:model}
The method of invariants, based on the irreducible representations of the space group at the $X$ point, yields an effective Bloch Hamiltonian
composed of a transverse part and a longitudinal part~\cite{jafari2026}. The longitudinal part is given by the following two-component (valleyor) Hamiltonian~\cite{jafari2026}:
\begin{equation}
    H_0 = \frac{p_z^2}{2 m_\ell} + \frac{\hbar k_1}{m_\ell}\tau_3 p_z,
    \label{eq:H0}
\end{equation}
where $p_z=-i\hbar\partial_z$, $m_\ell = 0.98m_e$ is the longitudinal effective mass, and $\hbar k_1$ is the momentum offset of the valley minima from $X$, which for silicon is given by $0.15\times 2\pi/a_0$ where $a_0=0.543$ nm is the size of the conventional unit cell of Si. At this level, it is worth noting that when the k.p theory is written with respect to the $X$ point, $k_1$ appears as a coupling constant in the effective Hamiltonian, and not as an oscillatory factor~\cite{Woods2024}. 
The Pauli matrix $\tau_3$ acts in the valley space $\{\ket{+},\ket{-}\}$ with $\tau_3\ket{\pm}=\pm\ket{\pm}$. The coupling constant in the second term is chosen such that the two minima are located at $\pm k_1=\pm 0.15\times 2\pi/a_0$ relative to the $X$ point, where $a_0$ is the length of the conventional Si unit cell. 
The wave function is confined to an infinite square well of length $L$, $\psi(\pm L/2) = 0$.

Motivated by the shear-strain coupling, which at the $X$ point takes the symmetry-allowed form $C_1\varepsilon_{xy}\tau_2$~\cite{Hensel1965,jafari2026}, we take the intervalley coupling to be $z$-independent,
\begin{equation}
    V = V_s \tau_2 ,
    \label{eq:V}
\end{equation}
where $V_s=C_1\varepsilon_{xy}$ is determined by the shear deformation potential $C_1$ and the shear strain component $\varepsilon_{xy}$. 
The reported values of $C_1$ vary considerably, ranging from $1.7$~eV obtained from tight-binding calculations~\cite{Woods2024} to theoretical estimates of $5.7$~eV~\cite{Hensel1965} and experimentally measured values $\approx 17.2\pm 0.8$~eV in Ref.~\cite{Laude1971}~\footnote{In this reference they use the symbol  $2{\cal E}_2^*$.}.
Putting these together, we obtain the full Hamiltonian in the valleyor basis:
\begin{equation}
    H = H_0 + V = \frac{p_z^2}{2 m_\ell} + \frac{\hbar k_1}{m_\ell}\tau_3 p_z + V_s\tau_2 .
    \label{eq:H}
\end{equation}
The above Hamiltonian follows solely from constructing an invariant under the space group of the $X$ point. In the following, we show that it is also consistent with the atomistic sp$^3$d$^5$s$^*$ tight-binding calculations of Woods \emph{et al.}~\cite{Woods2024}, who derived an effective-mass intervalley Hamiltonian for shear-strained Si/SiGe wells. To leading order, their strain-induced valley-off-diagonal term contains the term $H_v= -i\varepsilon_{xy}C_1e^{i2k_1z}$, with $C_1=1.73$~eV (cf. Eq.~(4) of~\cite{Woods2024}). 
To establish the connection between the valleyor representation introduced above and the conventional envelope-function representation used in Eq.~(4) of Ref.~\cite{Woods2024}, it is sufficient to perform the unitary transformation $U=e^{ik_1\tau_3z}$ on our valleyor Hamiltonian. This transformation shifts the origins of the left (right) parabola to $-k_1$ ($+k_1$), thereby yielding two copies of the standard parabolic kinetic-energy terms of the envelope-function approach on the diagonal:
\begin{eqnarray}
    H' &=& UHU^\dagger =
    \begin{pmatrix}
    \dfrac{p_z^2}{2m_\ell} & -iV_s e^{2ik_1z}\\[2mm]
    iV_s e^{-2ik_1z} & \dfrac{p_z^2}{2m_\ell}
    \end{pmatrix}-E_{\rm 1}\tau_0,
    \label{eq:Hprime}
\end{eqnarray}
where $E_{\rm 1}=\hbar^2k_1^2/(2m_\ell)$ is a constant that indicates energy of the $X$ point with respect to valley minima and $\tau_0$ is the 
unit $2\times 2$ matrix in the valley space. The above transformed Hamiltonian also correctly gives the off-diagonal coupling $-iV_se^{2ik_1z}$ appearing in the envelope function approach~\cite{Woods2024}. It is important to note that, although the envelope-function form~\eqref{eq:Hprime} in Ref.~\cite{Woods2024} is derived within a tight-binding framework, the factor $-i$ in the deformation potential $-iC_1$ appearing in their Eq. (4) corresponds precisely to the valley Pauli matrix $\tau_2$ obtained purely from our symmetry analysis of the space group at the $X$ point~\cite{jafari2026}. This provides an independent microscopic confirmation of the group-theoretical correspondence $\varepsilon_{xy}\leftrightarrow\tau_2$. Numerically, $C_1$ also fixes the scale of $V_s$. Taking $C_1\sim 10$~eV as a representative average of the reported values, a shear strain of $\varepsilon_{xy}\sim 0.1-0.5\%$ corresponds to $V_s\approx C_1\varepsilon_{xy}\approx10-50$~meV. The values used in the figures are chosen to be near this physically relevant value.

The disappearance of the oscillatory factors in Eq.~\eqref{eq:Hprime} of the envelope-function approach upon transforming to the 
valleyor basis, in which the Hamiltonian depends only on $k$ and parameters such as $k_1$ and $V_s$, makes the valleyor representation 
in Eq.~\eqref{eq:H} particularly powerful. 
The fact that the momentum offset $k_1$, rather than appearing as an oscillatory factor that complicates the Hamiltonian in Eq.~\eqref{eq:Hprime}, emerges as a \emph{coupling constant} in Eq.~\eqref{eq:H} suggests that the strength of $V_s$ should be assessed relative to the other natural energy scale of the Hamiltonian~\eqref{eq:H}, namely $E_1$.

The consistency between our approach based on writing an invariant Hamiltonian with respect to the space
group of the $X$ point~\cite{jafari2026} and the effective Hamiltonian projected from the tight-binding approach in Ref.~\cite{Woods2024}  provides further support for our approach, which is based on the symmetry analysis of the $X$ point from which the pair of valley states originates.
 The simplicity of the $X$-point representation enables us to construct an analytical variational wave function that captures the non-perturbative effects of both strain and wiggle-well potentials, thereby providing new insights into the effects of shear strain and the wiggle-well potential.
 In the envelope-function basis of Eq.~\eqref{eq:Hprime}, the Hamiltonian depends on both $k$ and $z$, requiring the wave equation to be solved numerically. 
We will show that the analytical variational wave functions for the two low-lying valley states obtained in the valleyor basis, 
when transformed to the envelope-function basis, agree very well with the numerical solutions of the traditional envelope method Eq.~\eqref{eq:Hprime}. 
This agreement establishes the valleyor basis at the $X$ point as a useful framework for developing an analytical understanding 
of the effects of various external perturbations on silicon spin qubits.
 The corresponding closed-form analytical treatment is presented in Appendix~\ref{app:derivation}. 

\begin{figure}[t]
    \centering
    \includegraphics[width=0.8\linewidth]{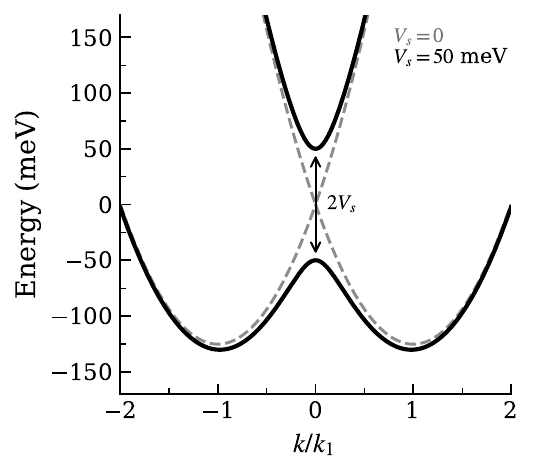}
    \caption{Dispersion near the $X$ point for $V_s=0$ (dashed) and
    $V_s=50$~meV (solid), Eq.~\eqref{eq:Epm}. The gap $2V_s$ opened at the
    former valley crossing is the plane-wave precursor of the finite-well
    valley splitting derived in Sec.~\ref{sec:ansatz}. As $V_s$ increases
    from zero, the minima of $E_-(k)$ move continuously toward the $X$ point.}
    \label{fig:valley_dispersion}
\end{figure}

The starting point for exploiting the simplicity of the valleyor basis is to consider plane-wave states $\psi(z) = e^{ikz}u(k)$ where $k$ is the Bloch wave vector along $z$ direction. In this representation, Eq.~\eqref{eq:H} is readily diagonalized in $k$-space, yielding
\begin{equation}
    E_\pm(k) = \frac{\hbar^2k^2}{2m_\ell} \pm
    \sqrt{\left(\frac{\hbar^2k_1k}{m_\ell}\right)^2 + V_s^2}.
    \label{eq:Epm}
\end{equation}
As shown in Fig.~\ref{fig:valley_dispersion}, $V_s$ opens a gap at the
previously degenerate valley crossing at the $X$ point. Because the gap
opens between the two branches that formerly belonged to the separate
valleys, both valleys are encoded into the lower branch $E_-(k)$
once $V_s\neq0$: the ground state no longer belongs to one valley or the
other, but is a hybrid of both.

As shown in Fig.~\ref{fig:valley_dispersion}, before the gap is opened by $V_s$, the conduction-band minima are located at $\pm k_1$. Once the crossing at the $X$ point is gapped by $V_s$, the minima shift toward the $X$ point, from $\pm k_1$ to $\pm k_{\rm min}$. Thus, the valley wave functions are determined by the combined effects of the parameters $k_1$ and $V_s$. Minimizing the lower branch $E_-(k)$ locates the shifted valley minima,
\begin{equation}
    k_{\rm min} = \sqrt{k_1^2 - \left(\frac{m_\ell V_s}{\hbar^2 k_1}\right)^2}
    =k_1\sqrt{1 - v_s^2},
    \label{eq:kmin}
\end{equation}
where the dimensionless strain energy scale
\begin{equation}
v_s=\left(\frac{V_s}{2E_1}\right)
\label{eq:ratio}
\end{equation}
naturally emerges. It is important to note that the energy scale $E_1$ associated with the $X$ point sets the natural scale for quantifying the effect of shear strain. This highlights the central role of the $X$ point, a topologically nontrivial point~\cite{Chamon2020}, in determining the valley splitting. In fact, as we will see from the central equation~\eqref{eq:splitting}, the valley splitting is governed by this dimensionless parameter.
Eq.~\eqref{eq:kmin} reduces to $k_{\rm min}=k_1$ for $V_s=0$ and moves toward the $X$ point as the intervalley coupling increases. The corresponding plane-wave ground state energy is
\begin{equation}
    E_0^{\rm pw} = E_-(k_{\rm min}) = -\frac{\hbar^2k_1^2}{2m_\ell} -
    \frac{m_\ell V_s^2}{2\hbar^2k_1^2}=-E_1(1+v_s^2).
\end{equation}
The plane-wave state is not an eigenstate of the finite well: the hard walls require an envelope with $\kappa=\pi/L$. Exact diagonalization shows $\Delta E\propto|\sin(k_{\rm min}L)|$, with nodes at $\varphi=k_{\rm min}L=n\pi$ and maxima at $\varphi=(2n+1)\pi/2$. We reproduce this boundary-condition--driven oscillation variationally below and demonstrate its agreement with the numerical solution of Eq.~\eqref{eq:Hprime}.

\section{Variational ansatz}
\label{sec:ansatz}
The valley-splitting physics is controlled by the mixing of the two initial valley
states $\ket{\pm}$ under the coupling term
$\frac{\hbar^2k_1k}{m_\ell}\tau_3 + V_s\tau_2$. Its (plane-wave) eigen-valleyor at arbitrary $k$ in the \textit{lower branch} of Fig.~\ref{fig:valley_dispersion} is
\begin{equation}
    \chi(k) = \frac{1}{\mathcal{C}(k)}
    \begin{pmatrix}
        V_s \\
        -i\left(\frac{\hbar^2k_1k}{m_\ell} +
        \sqrt{\left(\frac{\hbar^2k_1k}{m_\ell}\right)^2+V_s^2}\right)
    \end{pmatrix},
    \label{eq:chi}
\end{equation}
with $\mathcal{C}(k)$ the appropriate normalization, and we define the valleyors associated with the energy minima at $\pm k_{\rm min}$ as $\chi_\pm \equiv \chi(\pm k_{\rm min})$. They satisfy the normalization condition $|\chi_\pm|^2=1$ and the \emph{non-orthogonality}
\begin{equation}
    \chi_+^\dagger\chi_- = \chi_-^\dagger\chi_+ = \frac{m_\ell V_s}{\hbar^2k_1^2}=v_s ,
    \label{eq:chi_overlap}
\end{equation}
i.e., the two valley eigenstates are \emph{not} orthogonal once $V_s\neq0$;
this \emph{residual valleyor-overlap} controlled by the ratio $v_s$ of strain energy scale $V_s$ and the energy $E_1$ of the $X$ point in Eq.~\eqref{eq:ratio} is what ultimately produces the finite \emph{valley splitting}. 
Writing $\chi_\pm=(u_\pm,-iw_\pm)$ with $u_\pm,w_\pm$ real, the overlap in Eq.~\eqref{eq:chi_overlap} follows because the phase on the second component cancels between $\chi^\dagger$ and $\chi$ in the bilinear $u_+u_-+w_+w_-$, leaving a result that depends on $V_s$ only through the same combination that fixes $k_{\rm min}$ in Eq.~\eqref{eq:kmin}.

To satisfy the hard-wall boundary conditions we dress each plane wave with
the lowest allowed envelope function, $\cos(\kappa z)$ with $\kappa=\pi/L$, and form
the symmetric and antisymmetric combinations ($\pm$) induced by the symmetry with respect to the center of the well,
\begin{equation}
    \psi_\pm(z) = \frac{\mathcal{N}_\pm}{\sqrt{L}}\cos(\kappa z)
    \left(e^{ik_{\rm min}z}\chi_+ \pm e^{-ik_{\rm min}z}\chi_-\right)
    \label{eq:psipm}
\end{equation}
as variational wave functions for the lowest and first excited valley states with ${\cal N}_\pm$ being dimensionless normalization constants.
Note that the envelope $\cos(\kappa z)$ originates from the envelope-function basis in Eq.~\eqref{eq:Hprime}. Therefore, factors $e^{\pm ik_{\rm min}z}$ are required to shift the reference point to the $X$ point, where the valleyor basis is constructed.
Using Eq.~\eqref{eq:chi_overlap}, the probability density follows directly:
\begin{equation}
    |\psi_\pm(z)|^2 = \frac{2}{L}|\mathcal{N}_\pm|^2\cos^2(\kappa z)
    \left(1 \pm v_s\cos(2k_{\rm min}z)\right).
    \label{eq:density}
\end{equation}

\begin{figure}[t]
    \centering
    \includegraphics[width=0.95\linewidth]{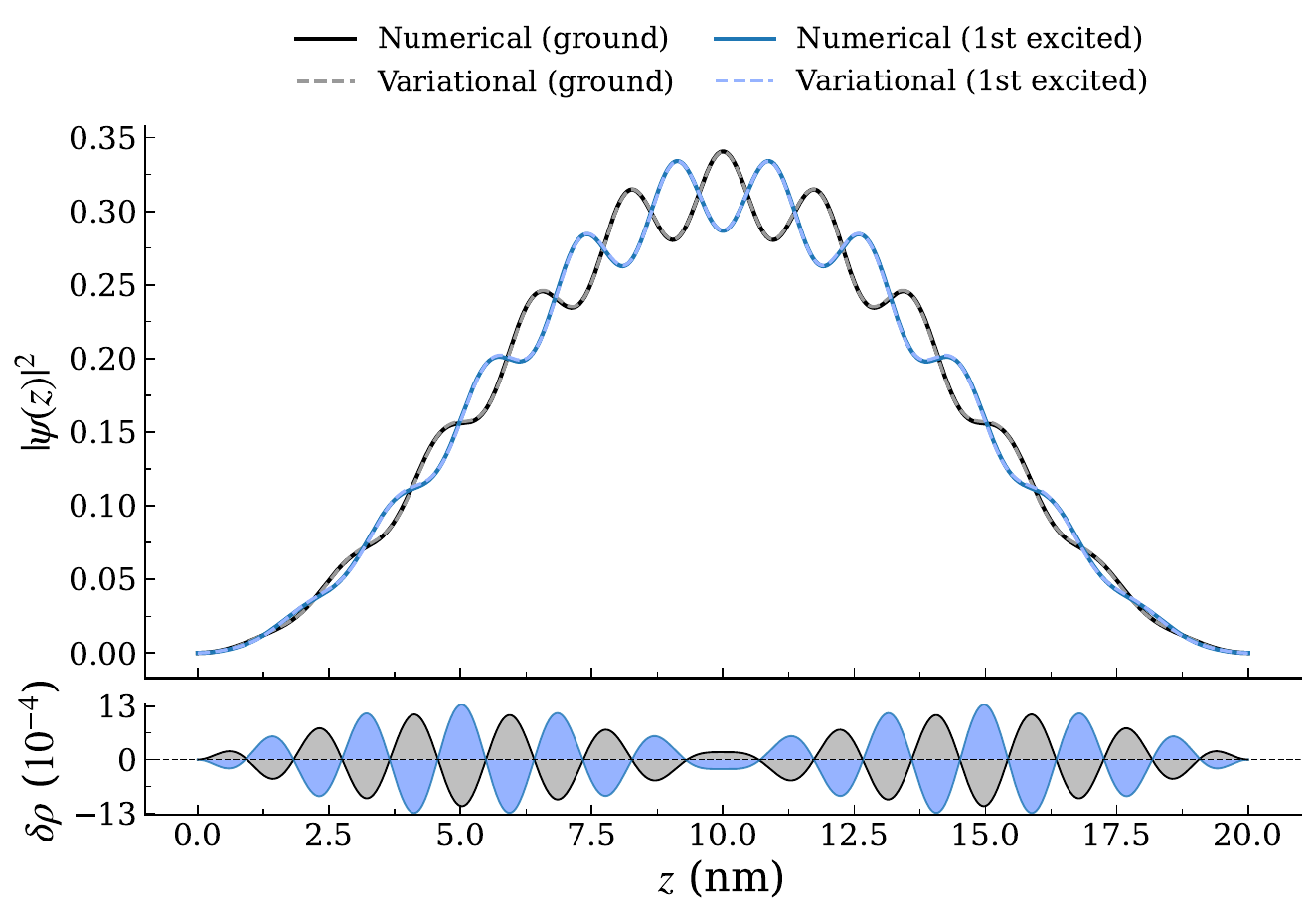}
    \caption{Probability density for the ground state ($\psi_+$) and first excited state ($\psi_-$), shown as gray and light-blue dashed lines, respectively, compared with the corresponding numerically exact results (solid black and blue lines) for $L=20$~nm and $V_s=20$~meV. The variational wave functions reproduce both the overall envelope and the fine-scale modulation of the numerical densities. The lower panel shows the residuals $\delta\rho=\rho_{\rm num}-\rho_{\rm var}$, which exhibits a beating pattern at the $10^{-4}$ level, below $\sim1\%$ of the density itself.}
    \label{fig:wavefunction_comparison}
\end{figure}

Fig.~\ref{fig:wavefunction_comparison} compares the wave function $\psi_+$ from Eq.~\eqref{eq:density} with our numerically exact ground-state density for $L=20$~nm, $V_s=20$~meV. Which of $\psi_+,\psi_-$ is the true ground
state is not fixed by the labeling in Eq.~\eqref{eq:psipm} -- it depends on
the sign of $E_+-E_-$ in Eq.~\eqref{eq:splitting}, i.e. on $L$, $V_s$, and
$k_1$ through $\sin(k_{\rm min}L)$. At
the parameters used throughout Figs.~\ref{fig:wavefunction_comparison}--\ref{fig:component_valleyor}
below, $L=20$~nm and $V_s=20$~meV, the ground state is $\psi_+$; we use this
same state in both figures. The agreement is very good, with only a small
deviation on the scale of $10^{-4}$ seen as residual beating pattern. This discrepancy arises from the fact that $\psi_\pm$ is not an exact eigenstate of $H$. The exact eigenstate has a broader peak around $k_{\rm min}$ in the Fourier spectrum, in contrast to the single-frequency variational ansatz.

The energy expectation values $E_\pm = \bra{\psi_\pm}H\ket{\psi_\pm}$ can be
evaluated in closed form; the full algebra is given in
Appendix~\ref{app:derivation}. The key simplification is that $\chi_\pm$
diagonalizes the valley-coupling term at $k=\pm k_{\rm min}$ exactly, so that acting
with $H$ on $f_\pm(z)\chi_\pm \equiv \cos(\kappa z)e^{\pm ik_{\rm min}z}\chi_\pm$ produces
a term proportional to $f_\pm\chi_\pm$ itself plus a residual term
proportional to $\sin(\kappa z)$ that mixes $\chi_+$ and $\chi_-$. It is this
mixing term, combined with the non-orthogonality in Eq.~\eqref{eq:chi_overlap},
that survives integration and yields the following analytical formula for the valley splitting:
\begin{eqnarray}
    &&\Delta E = |E_+-E_-| \nonumber\\
    &&=\frac{\kappa^2V_sk_{\rm min}}{Lk_1^2(\kappa^2-k_{\rm min}^2)}
    \left(|\mathcal{N}_+|^2+|\mathcal{N}_-|^2\right)|\sin\varphi|\nonumber\\
     &&=\frac{2 v_s\sqrt{E_\kappa \bar E_1}}{\pi (E_\kappa-\bar E_1)} E_\kappa
    \left(|\mathcal{N}_+|^2+|\mathcal{N}_-|^2\right)|\sin\varphi|   
    \label{eq:splitting}
\end{eqnarray}
where $\varphi=k_{\rm min}L$, $E_\kappa=\hbar^2\kappa^2/2m_\ell$ is the quantization energy scales of the quantum well, $E_1$ of the $X$ point with respect to valley minima, $\bar E_1=E_1(1-v_s^2)$ is renormalized energy of the $X$-point and the dimensionless strain energy parameter $v_s$ is given by Eq.~\eqref{eq:ratio}. 
The above equation encompasses all the relevant energy scales: $E_\kappa$ is a property of the heterostructure, $E_1$ is a characteristic 
energy scale of the $X$ point of the silicon universe in which the spin qubit resides, and $v_s$ quantifies the external influence of shear strain.

The normalization $\langle\psi_\pm|\psi_\pm\rangle=1$ gives,
\begin{eqnarray}
    |\mathcal{N}_\pm|^2 &&= \left(1 \mp\frac{2}{L}
    \frac{m_\ell V_s \kappa^2}{2\hbar^2k_1^2(k_{\rm min}^3-k_{\rm min}\kappa^2)}\sin\varphi\right)^{-1}\nonumber\\
    &&= \left(1 \mp\frac{V_sE_\kappa}{2\pi E_1\bar E_1}\frac{\sqrt{E_\kappa \bar E_1}}{\bar E_1-E_\kappa}\sin\varphi\right)^{-1}.
    \label{eq:norm}
\end{eqnarray}
where again as in Eq.~\eqref{eq:splitting} $\bar E_1=E_1(1-v_s^2)$. 
Equation~\eqref{eq:splitting} is the central analytical result of this
paper: it reproduces the $|\sin(k_{\rm min}L)|$ dependence seen numerically, with a
prefactor determined entirely by the relevant energy scales $E_\kappa$ of the quantum well, $E_1$ of the Si band structure
and $v_s$ of the shear strain. The derivation in Appendix~\ref{app:derivation} depends on $\chi_\pm$ only through the bilinears $u_+u_-+w_+w_-$ and $|u_\pm|^2-|w_\pm|^2$ that enter Eq.~\eqref{eq:chi_overlap} and the $\tau_3$ expectation values used there.

\begin{figure}[tb]
    \centering
    \includegraphics[width=0.95\linewidth]{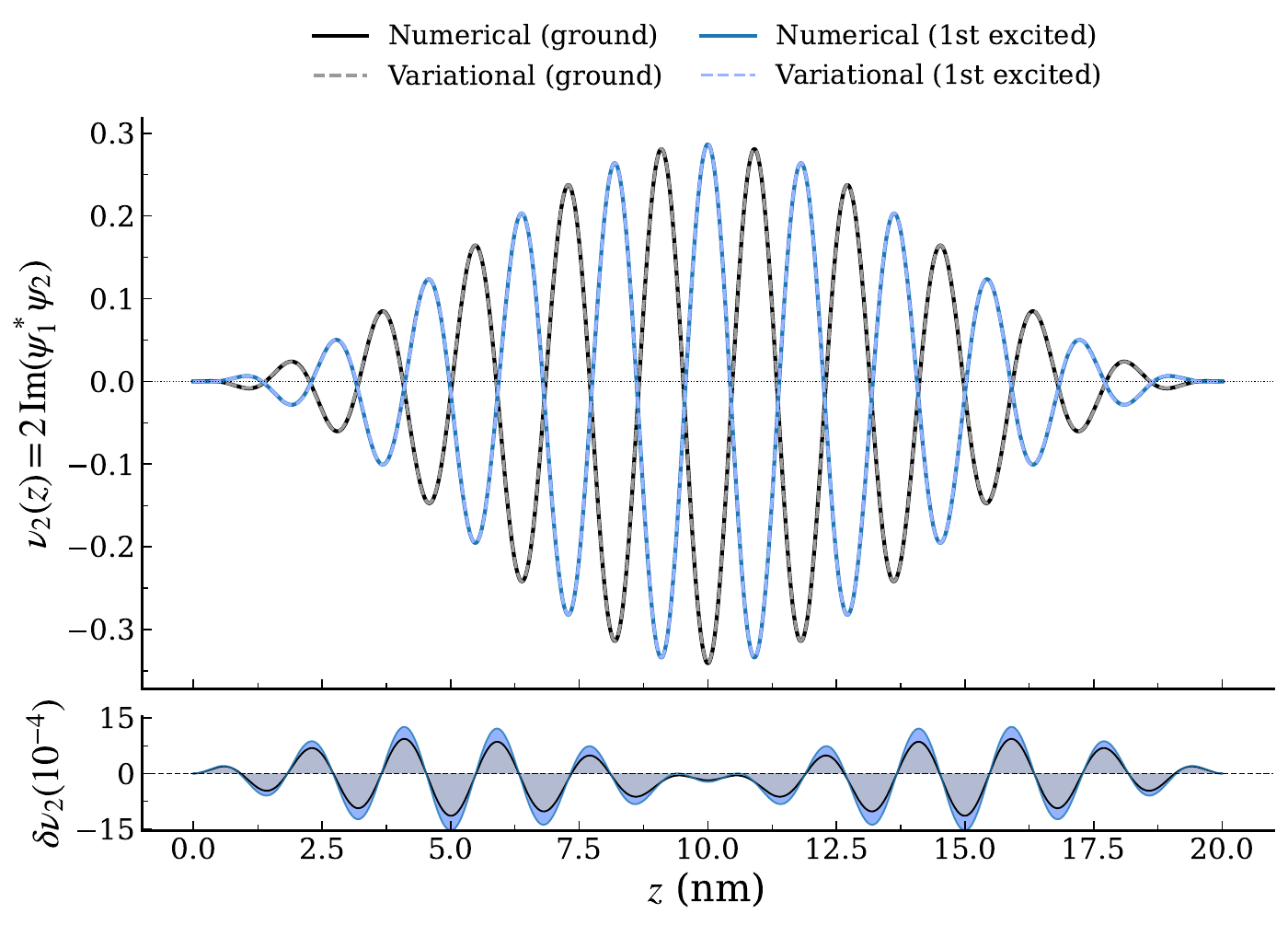}
    \caption{Intervalley interference $\nu_2(z) = 2\,\mathrm{Im}(\psi_1^*\psi_2)$,
    Eq.~\eqref{eq:tau2_density}, for the ground state ($\psi_+$ (black) and excited valley state $\psi_-$ (blue); see
    Sec.~\ref{sec:ansatz}): numerical (solid) vs. variational (dashed)
    for the same parameters as Fig.~\ref{fig:wavefunction_comparison}.
    As can be seen the two states are characterized by opposite valley polarization densities. 
    Lower panel: residual $\delta \nu_2 = \nu_{2,\rm num}-\nu_{2,\rm var}$.}
    \label{fig:component_valleyor}
\end{figure}

\section{Valley density of the ansatz wave function}
\label{sec:valleyor}
The two-component wave function also defines a local valley density,
\begin{equation}
    \nu_i(z) \equiv \psi^\dagger(z)\,\tau_i\,\psi(z), \qquad i=1,2,3,
    \label{eq:valleyor_def}
\end{equation}
so that $\nu_3(z) = |\psi_1(z)|^2-|\psi_2(z)|^2$ is the local $\tau_3$-type valley
polarization density and $\nu_1(z) = 2\,\mathrm{Re}[\psi_1^*(z)\psi_2(z)]$, $\nu_2(z) = 2\,\mathrm{Im}[\psi_1^*(z)\psi_2(z)]$ are the
local $\tau_1$- and $\tau_2$-type valley polarization densities formed by the "interference" between the $\ket{+}$ and $\ket{-}$ valley components of the wave function at every point $z$.
For a real, symmetric confining potential, $T=\tau_1\mathcal{K}$ (with $\mathcal{K}$ complex conjugation) commutes with $H$. A non-degenerate eigenstate can therefore be chosen to satisfy $T\psi=\psi$, implying $|\psi_1|=|\psi_2|$ and hence $\nu_3(z)=0$. Thus, the two valleys have equal local populations, while the intervalley interference is carried by $\nu_1$ and $\nu_2$. For our ansatz these valley polarization densities are given by:
\begin{eqnarray}
    \nu_2(z) &&= -\frac{\cos^2(\kappa z)}{\mathcal{N}}
    \left[v_s \pm \cos(2k_{\rm min}z)\right],
    \label{eq:tau2_density}\\
    \nu_1(z) &&= \pm\frac{\cos^2(\kappa z)}{\mathcal{N}}\,\sqrt{1-v_s^2}\,
    \sin(2k_{\rm min}z),
    \label{eq:tau1_density}
\end{eqnarray}
where $\mathcal{N}$ is fixed by $\int \rho(z)\,dz=1$ as in Eq.~\eqref{eq:norm}, and the $\pm$ corresponds to the states $\psi_\pm$. Since $\nu_3(z)=0$, the identity
$\sum_i\nu_i^2=\nu_1^2+\nu_2^2=\rho^2$
also holds. 

Figure~\ref{fig:component_valleyor} compares the nontrivial valleyor
component, the intervalley interference $\nu_2(z)$ of
Eq.~\eqref{eq:tau2_density}, to the same exact diagonalization for the
ground and excited valley states $\psi_+$ and $\psi_-$; (see Sec.~\ref{sec:ansatz} and
Fig.~\ref{fig:wavefunction_comparison}). As can be seen, the solid (numerical) and dashed (analytical) curves for each state (color) are in excellent agreement. The deviation between them, shown in the bottom panel of the figure, reproduces the characteristic beating pattern already observed in Fig.~\ref{fig:wavefunction_comparison}, with a residual on the order of $\sim\!10^{-4}$. Furthermore, the standing valley density wave pattern for $\psi_+$ and $\psi_-$ has opposite polarization at every point $z$. 

The total valley polarization of the ansatz states will be given by integrating the above densities over the $z$:
\begin{equation}
   \nu_i=\int dz \nu_i(z).
\end{equation}
This only produces a non-zero $\nu_2$ as expected form of coupling $V_s\tau_2$.

\section{Benchmarking against numerically exact diagonalization}
\label{sec:benchmark}

\begin{figure}[t]
    \centering
    \includegraphics[width=0.95\linewidth]{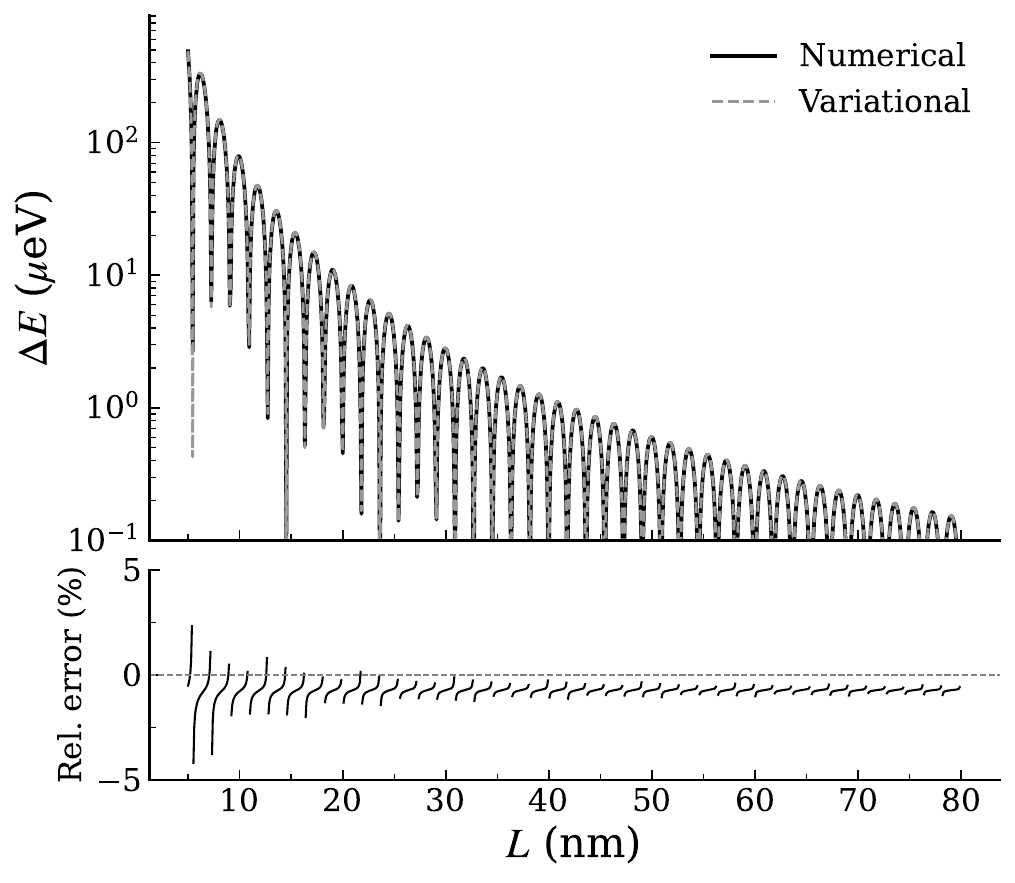}
    \caption{Valley splitting as a function of the heterostructure size $L$.
The variational prediction from Eq.~\eqref{eq:splitting}, shown as a dashed gray line, reproduces both the oscillatory node structure and the overall magnitude of the numerically exact splitting (solid black) as $L$ varies from 5 to 80~nm at fixed $V_s=20$~meV. Lower panel: relative error (with splitting nodes masked; see text). The error remains essentially flat at the $\sim\!1\%$ level across the entire range.}
    \label{fig:splitting_vs_L}
\end{figure}

\begin{figure}
    \centering
    \includegraphics[width=0.95\linewidth]{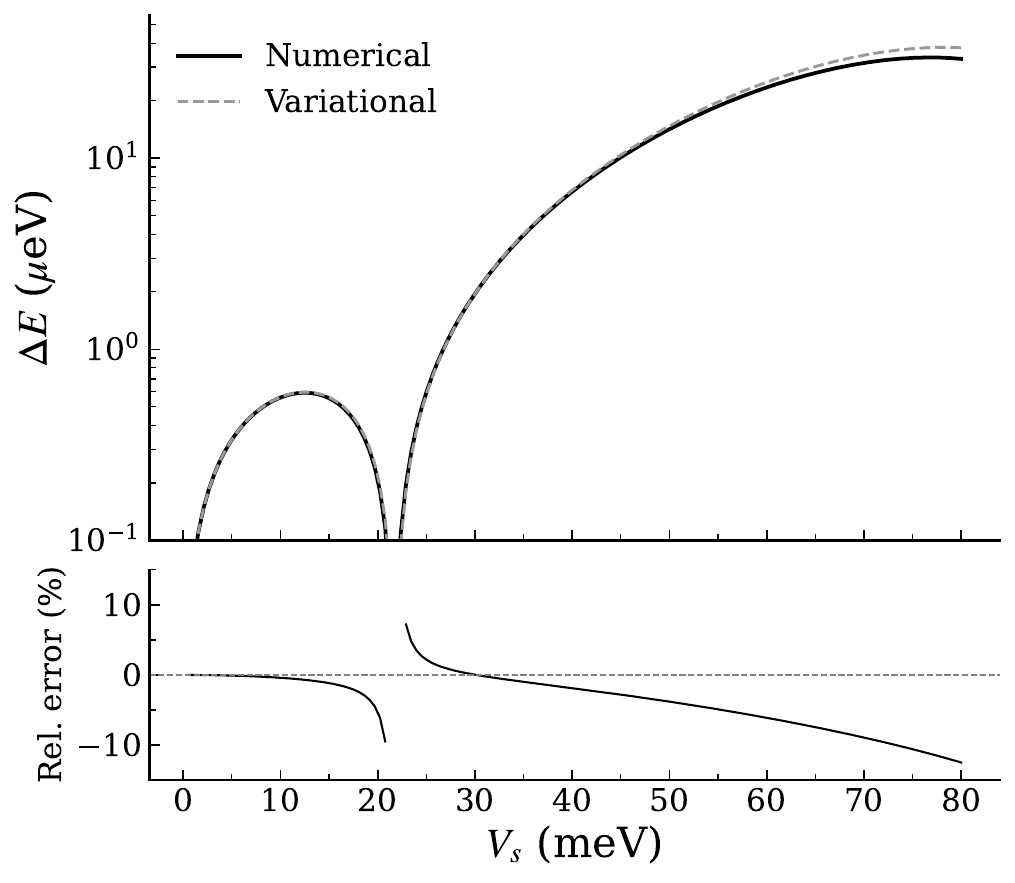}
    \caption{Unlike the well-width sweep of Fig.~\ref{fig:splitting_vs_L},
    sweeping the coupling strength $V_s$ at fixed $L=20$~nm exposes a
    systematic breakdown of the ansatz: numerics (solid black) and
    Eq.~\eqref{eq:splitting} (dashed gray) agree to $\lesssim\!1\%$ at weak
    coupling, but the theory increasingly overestimates $\Delta E$ as $V_s$
    grows, reaching $\sim\!10\%$ by $V_s\approx80$~meV (lower panel;
    splitting node masked, see text). }
    \label{fig:splitting_vs_V1}
\end{figure}

The comparisons above are for a single, representative choice of $L$ and
$V_s$. To establish the broad range of validity of the analytical result Eq.~\eqref{eq:splitting}, we
compare it against exact diagonalization over two broad sweeps: $\Delta
E(L)$ at fixed $V_s=20$~meV (Fig.~\ref{fig:splitting_vs_L}), and $\Delta
E(V_s)$ at fixed $L=20$~nm (Fig.~\ref{fig:splitting_vs_V1}). In both cases
the lower panel shows the relative error $100\times(\Delta E_{\rm
num}-\Delta E_{\rm var})/\Delta E_{\rm var}$, with points near a splitting node masked out to avoid divergences. As established in Sec.~\ref{sec:model}, both $\Delta E$ and the exact diagonalization it is benchmarked against depend on the coupling only through its magnitude $V_s$, regardless of which transverse valleyor matrix mediates it; only the interference structure of Sec.~\ref{sec:valleyor} is sensitive to that choice.

As can be seen in Fig.~\ref{fig:splitting_vs_L}, the variational result stays within $\sim1\%$ of exact diagonalization over
$L=5$--$80$~nm at $V_s=20$~meV, including the oscillatory node structure and
the overall scale of the splitting. On the other hand as can be seen in Fig.~\ref{fig:splitting_vs_V1}, at fixed $L=20$~nm, the error likewise
remains $\lesssim1\%$ for realistic $V_s=10$--$30$~meV but grows systematically
at stronger coupling, reaching $\sim10\%$ near $80$~meV. The deviation arises
from the rigid single-harmonic envelope in Eq.~\eqref{eq:psipm}: the exact
state can lower its energy by mixing higher box harmonics. This relaxation
becomes more important with increasing $V_s$, causing the variational result
to overestimate $\Delta E$. Over the parameter range studied, Eq.~\eqref{eq:splitting}
therefore acts as an empirical upper bound, although we do not prove this
analytically. 

Setting aside the quantitative accuracy, our analytical result provides a natural explanation for the oscillatory behavior $|\sin\varphi|$, where $\varphi=k_{\rm min}L$ combines $k_{\rm min}$, which is determined by $V_s$ for fixed $k_1$ as in Eq.~\eqref{eq:kmin}, and the well width $L$. Therefore, the nodes and antinodes observed as a function of $L$ in Fig.~\ref{fig:splitting_vs_L} are also expected as a function of $V_s$ in Fig.~\ref{fig:splitting_vs_V1}, since varying $V_s$ changes $k_{\rm min}$.
This observation is particularly important because, for a given device, the well width $L$ is fixed and cannot be varied, whereas $V_s$ can potentially be tuned by varying the shear strain $\varepsilon_{xy}$.

Therefore, for fixed $L$, the node condition $\varphi=k_{\rm min}(V_s)L=n\pi$ also manifests itself in the shear-strain dependence 
of the valley splitting, appearing as a canyon-shaped suppression of the valley splitting in Fig.~\ref{fig:splitting_vs_V1}. 
Importantly, further increasing the strain beyond this minimum leads to a substantial enhancement of the valley splitting, 
which can reach values an order of magnitude larger on the left side of the canyon.

\begin{figure}[t]
\centering
\includegraphics[width=0.95\linewidth]{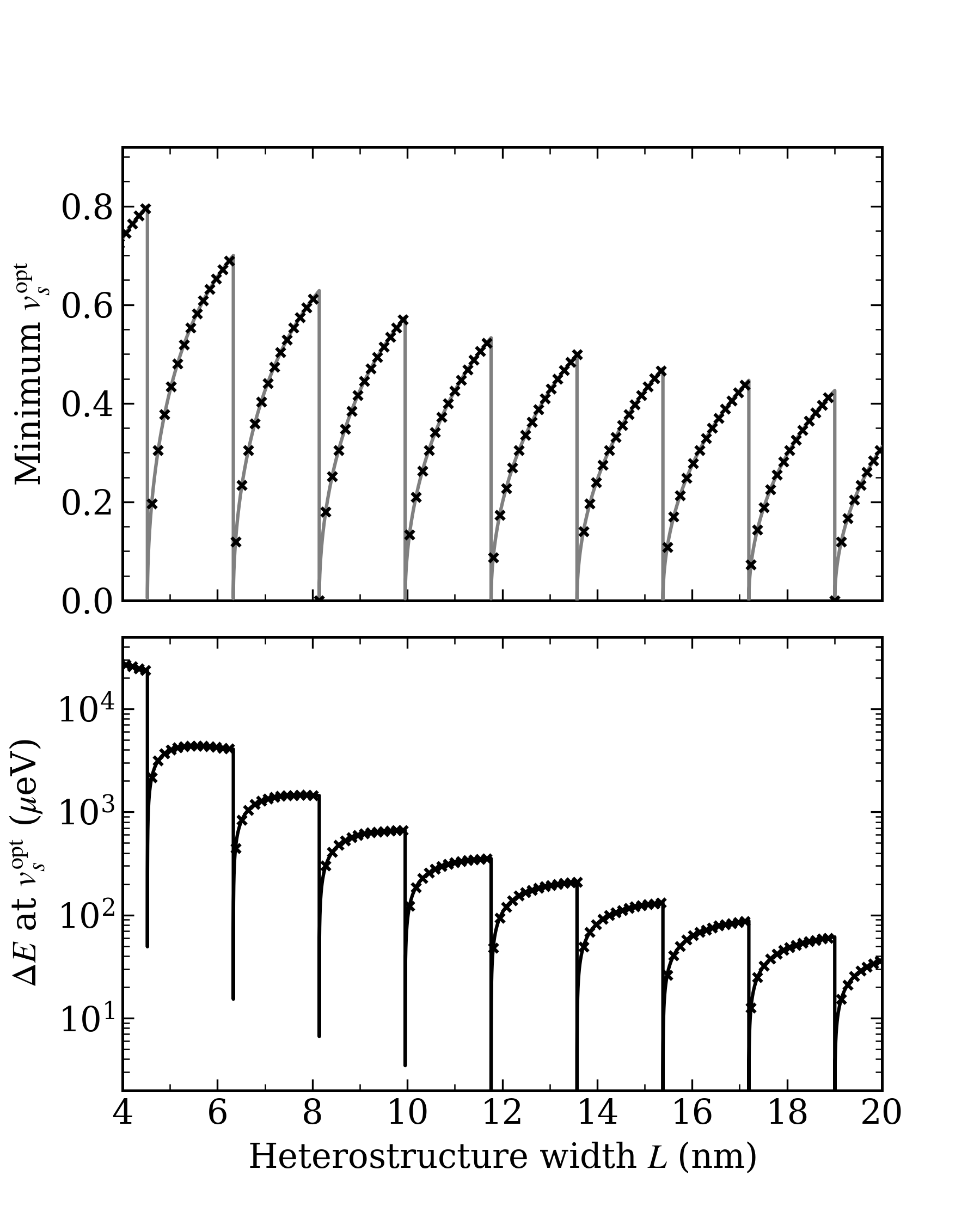}
\caption{Top: the lowest value of the dimensionless shear strain energy scale $v_s^{\rm opt}$ at antinodes of the $\sin(\varphi)$ oscillation.  Bottom: the valley splitting $\Delta E$ for $v_s^{\rm opt}$ at each $L$. Crosses denote heterostructures with $L = ma_0/4$ for integer $m$ that corresponds to the
number of monolayers.}
\label{fig:vs_opt}
\end{figure}

Our analytical formula allows us to estimate the optimal shear strain for a given heterostructure width $L$ as follows. The periodic dependence on shear strain discussed above suggests a simple rule of thumb for choosing the optimal strain values. One simply needs to avoid the nodes at $\varphi=n\pi$. 
Therefore one can choose the strain such that $k_{\rm min}L=(n-1/2)\pi$, corresponding to the $n$th maximum of $|\sin\varphi|$. With this choice, one obtains
\begin{eqnarray}
    v_s^{\rm opt}&&=\sqrt{1-(n-1/2)^2\frac{E_\kappa}{E_1}}=\sqrt{1-\left(\frac{\pi(n-1/2)}{k_1L}\right)^2}\nonumber\\
    &&=\sqrt{1-\left(\frac{(n-1/2)a_0}{0.3L}\right)^2}
    \label{eq:vs_opt}
\end{eqnarray}
where we have used $k_1=0.15\times 2\pi/a_0$ with $a_0=0.543$ nm the lattice constant of Si. 
The dimensionless parameter $v_s^{\rm opt}$, when multiplied by $E_1\approx 120~$meV, gives the corresponding energy scale $V_s$. 
Equation~\eqref{eq:vs_opt} is remarkable in that it identifies the shear-strain values corresponding to the antinodes of 
the energy scale $V_s$ solely in terms of the heterostructure width $L$. To attain the smallest possible value of $v_s^{\rm opt}$, 
one must choose the largest possible integer $n_{\rm max}$, which is given by:
\begin{equation}
    n_{\mathrm{max}} = \left\lfloor \frac{0.3 L}{a_0} + \frac{1}{2} \right\rfloor
    \label{eq:nopt}
\end{equation}
For a handful of $L$ values from $4-20~$nm, the above formula gives the plot in Fig.~\ref{fig:vs_opt}.
The results shown in this figure indicate that the optimal strain must be tuned according to the width $L$ of the heterostructure.
Note that the value of valley splitting can be obtained by feeding these anti-node values of the dimensionless strain parameter $v_s$ to Eq.~\eqref{eq:splitting}. 

\section{Wiggle-well model}
\label{sec:wiggle}
We now show that our valleyor representation provides useful insight into two distinct regimes of the wiggle-well potential.
To model the WW, we consider a Ge concentration profile that is periodic
along $z$ with wavelength $\lambda = 2\pi/k_w$. This potential couples the valley states; the corresponding
effective Hamiltonian in the valleyor basis at the $X$  point becomes
\begin{equation}
H_{\rm wiggle}
=
\frac{p_z^2}{2m_\ell}
+\frac{\hbar k_1}{m_\ell}\tau_3p_z
+V_w\cos(k_wz)\tau_1 ,
\label{eq:Hwiggle}
\end{equation}
with hard-wall boundaries $\psi(\pm L/2)=0$. A scalar wiggle, i.e. $V_w\cos(k_wz)\tau_0$ does not introduce a valley splitting.
This equation becomes equivalent to Eq.~(4) of Ref.~\cite{Woods2024} upon applying the unitary transformation discussed in Eq.~\eqref{eq:Hprime}, 
which shifts the origin of the momentum axis to $\pm k_1$ to conform with their envelope-function approach. Once again, the advantage of 
working in the valleyor basis is that it eliminates most of the redundant oscillatory factors, resulting in a simpler representation 
that is more amenable to analytical treatment as shown below. 
For the parameter $V_w$ quantifying the amplitude of the wiggle, we use a representative value of $V_w=20$ meV~\cite{Woods2024}. This value can also be estimated from the following reasoning: Typical values of the conduction band offset between Si and Ge lie in the range of $20$–$80$ meV~\cite{People1986,Yu1992}. For a Ge concentration of $30\%$, this is effectively reduced by a factor of $0.3$, yielding 
a rough estimated range of $10$–$30$ meV for $V_w$.

In Eq.~\eqref{eq:Hwiggle}, $k_w$ is a model parameter and can be tuned to the intervalley resonance at
\begin{equation}
k_w=2k_1 ~~~\mbox{resonance condition}.
\end{equation}
To make this resonance explicit, we apply the same unitary transformation
$U=e^{ik_1\tau_3z}$ used in Eq.~\eqref{eq:Hprime}. In the transformed frame, the two Fourier components of the modulation oscillate with wavevectors
$2k_1-k_w$ and $2k_1+k_w$. Thus,
\begin{equation}
\cos(k_wz)\tau_1\nonumber
\longrightarrow
\frac{1}{2}
\left[
e^{i(2k_1-k_w)z}\tau_+
+e^{i(2k_1+k_w)z}\tau_+
+\mathrm{h.c.}
\right].
\end{equation}
Near $k_w=2k_1$, the first term is slowly varying while the second oscillates
rapidly on the scale of the well. Since $k_1L\approx35$ at $L=20$~nm, the rapidly oscillating contribution is negligible in the low-energy matrix elements. 
We therefore make a rotating-wave approximation, obtaining
\begin{equation}
H_{\rm res}
=
\frac{p_z^2}{2m_\ell}
+\frac{\hbar q}{m_\ell}\tau_3p_z
+\frac{V_w}{2}\tau_1,
\qquad
q=k_1-\frac{k_w}{2}.
\label{eq:Hres_valleyor}
\end{equation}
where $q$ denotes half the detuning wave vector, which appears as a coupling constant in the valleyor representation rather than as an oscillatory term in the envelope-function Hamiltonian. 
Keeping the discarded rapidly oscillating term changes the calculated
splitting by less than $10^{-3}$~meV over the parameter range considered here.
Equation~\eqref{eq:Hres_valleyor} has the same structure as the constant-coupling
Hamiltonian of Eq.~\eqref{eq:H}, with the replacements
$k_1\rightarrow q$ and $V_s\rightarrow V_w/2$. 
The essential difference is that $q$ does not originate from the underlying band structure of silicon but 
is instead imposed by the external potential. It therefore plays the role of an ``effective valley displacement'' 
that vanishes at the physical resonance $k_w=2k_1$.

\subsection{Two regimes}
The lower dispersion branch in the rotating-wave approximation is given by
\begin{equation}
E_-(k)
=
\frac{\hbar^2k^2}{2m_\ell}
-\frac{1}{2}
\sqrt{
V_w^2+
\left(\frac{2\hbar^2qk}{m_\ell}\right)^2
},
\end{equation}
with minima at
\begin{equation}
k_{w\rm,min}=\pm
\sqrt{
q^2-
\left(
\frac{m_\ell V_w}{2\hbar^2q}
\right)^2
}.
\label{eq:k0_wiggle}
\end{equation}
For
\begin{equation}
|2k_1-k_w|= |2q|
>
\sqrt{\frac{2m_\ell V_w}{\hbar^2}},
\label{eq:wiggle_double_min}
\end{equation}
the lower branch has two minima at $\pm k_{w\rm,min}$. The low-energy states can
then still be regarded as two valley-like states, and the same logic as before can be followed.

At the resonance, as $k_w$ approaches $2k_1$ ($q\to 0$), the two minima move toward one another and
eventually merge at $k_{w\rm,min}=0$. At that point the relevant low-energy states are the lowest
two orbital states of a single hybridized $\tau_1$ branch. 
The WW potential therefore provides a direct crossover between valley-dominated physics far from resonance 
and conventional orbital quantization near resonance. This insight is made possible by the natural transparency 
and simplicity of the valleyor basis.

\subsection{Far from resonance: valleyor description}
When $k_{w\rm,min}$ is real, the lower-branch valleyors
$\chi_\pm=\chi(\pm k_{w\rm,min})$ are nonorthogonal, just as in the constant-coupling
problem. The variational construction of Eq.~\eqref{eq:psipm} therefore carries
over under
$k_{\rm min}\rightarrow k_{w\rm,min}$, $k_1\rightarrow q$, and
$V_s\rightarrow V_w/2$:
\begin{equation}
\Psi_\pm(z)
=
\frac{\mathcal N_\pm}{\sqrt L}
\cos(\kappa z)
\left[
e^{ik_{w\rm,min}z}\chi_+
\pm
e^{-ik_{w\rm,min}z}\chi_-
\right],
\label{eq:wiggle_valley_ansatz}
\end{equation}
where $\kappa =\frac{\pi}{L}$. Repeating the calculation leading to Eq.~\eqref{eq:splitting} gives
\begin{equation}
\Delta E^{\rm w}
=
\frac{
\kappa^2V_wk_{w\rm,min}
}{
2Lq^2(\kappa^2-k_{w\rm,min}^2)
}
\left(
|\mathcal N_+|^2+|\mathcal N_-|^2
\right)
|\sin(k_{w\rm,min}L)| ,
\label{eq:wiggle_valley_splitting}
\end{equation}
where $|\mathcal N_\pm|^2$ are given by the direct counterparts of Eq.~\eqref{eq:norm}.

\subsection{Near resonance: orbital description}
Close to resonance the coupling constant set by the detuning $q$ away from $2k_1$ accompanying the $\tau_3$ term in Eq.~\eqref{eq:Hres_valleyor} vanishes,
and therefore it becomes more natural to work directly in the particle-in-a-box basis
\begin{equation}
\phi_n(z)=\sqrt{\frac{2}{L}}\sin\left(n \kappa z\right),\quad \kappa=\frac{\pi}{L}.
\end{equation}
In the rotating-wave approximation, the intervalley coupling between orbitals
$m$ and $n$ is
\begin{align}
W_{mn}(k_w)
&=
\frac{V_w}{2}C_{mn}(2k_1-k_w),\\
C_{mn}(k)&=
\int_0^L\phi_m(z)\phi_n(z)e^{ikz}dz .
\end{align}

At exact resonance, $k_w=2k_1$, the coupling becomes spatially uniform in the
transformed frame, so that $C_{mn}(0)=\delta_{mn}$. The two orbitals therefore
decouple, and the splitting reduces to the ordinary orbital level spacing,
\begin{equation}
\Delta E = \frac{3\kappa^2\hbar^2}{2m_\ell}=3E_\kappa,
\label{eq:deltaE_max}
\end{equation}
where, as before, $E_\kappa$ denotes the natural energy scale of the quantum well.
Away from resonance, the orbital matrix elements decrease according to the
finite-well form factor $C_{mn}$, producing the diffraction-like sidelobes seen
in the numerical spectrum of Fig.~\ref{fig:splitting_vs_kw}. Enlarging the orbital basis converges this
description onto the exact numerical diagonalization of $H_{\rm wiggle}$.

\subsection{Strong-coupling limit: Asymptotic freedom for spin qubits}
The near-resonance regime admits a particularly simple limiting form when $V_w$ is large compared with the orbital level spacing, a condition that is physically relevant for $L=20$ nm.
Contrary to the intuition that a dominant WW potential amplitude $V_w$ should lead to substantial deviations 
from the empty-box limit, we find that the WW amplitude contributes only a subleading correction to the empty-box properties in this regime.
To show this, we write Eq.~\eqref{eq:Hres_valleyor} as
\begin{equation}
    H_{\rm res}=H_0+H_q,
    \quad
    H_0=
    \frac{p_z^2}{2m_\ell}
    +\frac{V_w}{2}\tau_1,
    \quad
    H_q=
    \frac{\hbar q}{m_\ell}p_z\tau_3 .
\end{equation}
The eigenstates of $H_0$ are the box orbitals
$\ket{n}$ multiplied by $\tau_1$ eigenstates $\ket{\pm}$:
\begin{equation}
    \ket{n,\pm}=\ket{n}\ket{\pm},
    \qquad
    E_{n,\pm}^{(0)}
    =
    E_n^{\rm box}\pm\frac{V_w}{2}.
\end{equation}
Since $\tau_3$ anticommutes with $\tau_1$, the detuned wiggle perturbation $H_q$ couples 
the two $\tau_1$ branches while having no matrix elements within either branch. 
Consequently, the first-order correction to the lower branch vanishes, while the second-order correction is
\begin{equation}
    \delta E_n^{(2)}
    =
    \sum_m
    \frac{
        \left|
        \bra{m,+}H_q\ket{n,-}
        \right|^2
    }{
        E_{n,-}^{(0)}-E_{m,+}^{(0)}
    } .
    \label{eq:wiggle_second_order}
\end{equation}
For $V_w$ much larger than the orbital level spacing, the denominator can be
approximated by
\begin{equation}
    E_{n,-}^{(0)}-E_{m,+}^{(0)}
    =
    E_n^{\rm box}-E_m^{\rm box}-V_w
    \simeq -V_w .
\end{equation}
Using
$\langle +|\tau_3|-\rangle=1$ (up to an irrelevant phase), Eq.~\eqref{eq:wiggle_second_order}
then becomes
\begin{align}
    \delta E_n^{(2)}
    & \simeq
    -\frac{\hbar^2q^2}{m_\ell^2V_w}
    \sum_m
    \left|\bra{m}p_z\ket{n}\right|^2 \nonumber \\
    & = -\frac{\hbar^2q^2}{m_\ell^2V_w}\bra{n}
    p_z^2\ket{n}
\end{align}
For the box eigenstates,
\begin{equation}
    \bra{n}p_z^2\ket{n}
    =
    \hbar^2\kappa^2n^2,
    \qquad
    E_n^{\rm box}
    =
    \frac{\hbar^2\kappa^2n^2}{2m_\ell}=n^2 E_\kappa.
\end{equation}
Thus the lower-branch energies are
\begin{equation}
    E_n
    \simeq
    E_n^{\rm box} -\frac{\hbar^4\kappa^2n^2q^2} {m_\ell^2V_w}
    = E_n^{\rm box} -\frac{4n^2 E_\kappa E_q} {V_w}.
    \label{eq:wiggle_strong_En}
\end{equation}
where, instead of $\hbar^2k_w^2/(2m_\ell)$, the energy scale $E_q=\hbar^2 q^2/(2m_\ell)$ associated with the detuning wave vector $q$ emerges. 

\begin{figure}[t]
\centering
\includegraphics[width=0.95\linewidth]{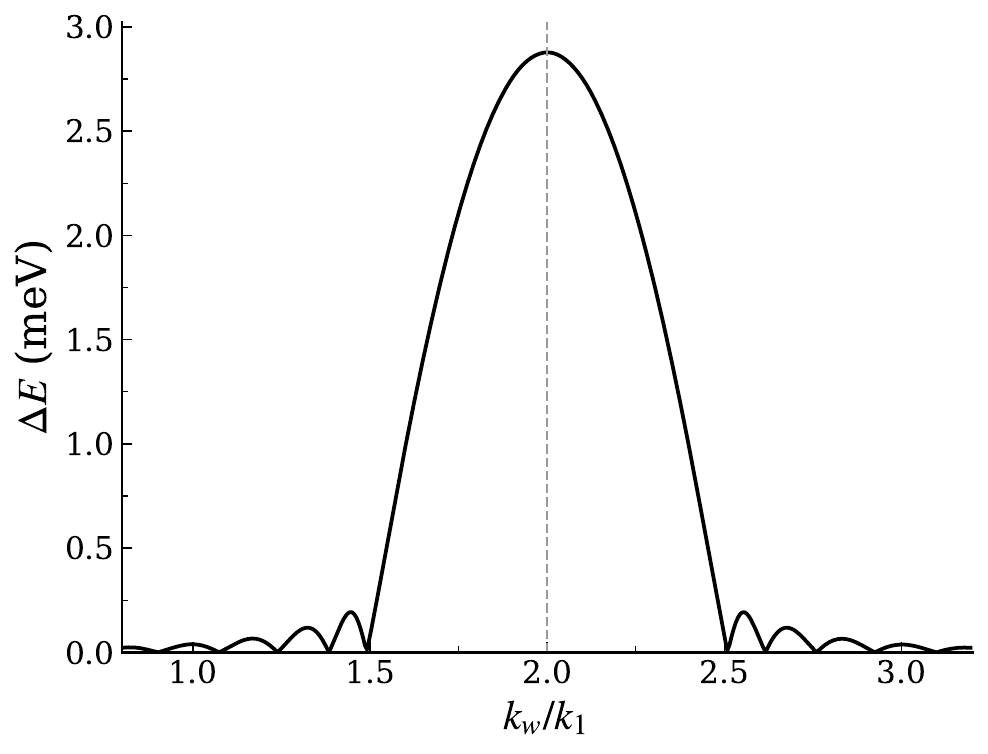}
\caption{Valley splitting of the wiggle-well Hamiltonian,
Eq.~\eqref{eq:Hwiggle}, as $k_w$ is scanned across $k_w=2k_1$, at fixed
$L=20$~nm and $V_w=20$~meV. The splitting peaks at $k_w=2k_1$
($\Delta E_{\rm max}\approx2.872$~meV, Eq.~\eqref{eq:deltaE_max} predicts 2.877~meV).}
\label{fig:splitting_vs_kw}
\end{figure}

Taking the difference between the $n=2$ and $n=1$ levels gives
\begin{equation}
    \Delta E
    \simeq
    \frac{3\kappa^2\hbar^2}{2m_\ell}
    \left[
        1-\frac{2\hbar^2q^2}{m_\ell V_w}
    \right]
    =3E_\kappa
    \left[
        1-\frac{2\hbar^2q^2}{m_\ell V_w}
    \right].
    \label{eq:wiggle_strongV}
\end{equation}

This derivation also makes the physical origin of the correction transparent:
the detuning term virtually admixes the upper $\tau_1$ branch, and the sum over
all intermediate orbitals by completeness reduces to the expectation value
of $p_z^2$. Thus the leading correction is proportional to $q^2/V_w$ explaining the central dome in Fig.~\ref{fig:splitting_vs_kw}.
At exact resonance, $q=0$, the virtual admixture disappears and the splitting
is simply the orbital level spacing. 
The box-quantization dominated valley splitting for $L=20~$nm at exact $2k_1$ resonance in Fig.~\ref{fig:splitting_vs_kw} is  $2.87~$meV. 
Increasing $V_w$ suppresses the virtual
correction and therefore makes the resonant splitting less sensitive to detuning.
The regime in which the energy scale \(V_w\) dominates over the energy scale \(E_q\) associated with the 
detuning-induced coupling \(q\) in Eq.~\eqref{eq:Hres_valleyor} can be referred to as ``asymptotic freedom'' 
for spin qubits.~\footnote{The original term refers to the weakening of interactions between particles at higher 
energies in certain gauge theories. In our case, as the wiggle-well potential depth \(V_w\) increases, the spin qubit 
increasingly behaves like a free particle in a box, with the valley splitting becoming dominated by the box-quantization energy scale.}
The practical implication of this observation is that precise tuning of the wiggle wavelength to the ($2k_1$) resonance 
is not essential, since some degree of detuning is inevitable in an experiment. The key requirement is instead to increase 
the wiggle-well amplitude ($V_w$) by suitable processing~\cite{Gradwohl2025},
thereby entering the strong-$V_w$ regime in which the valley splitting is governed predominantly by the box-quantization energy scale.

\section{Finite-depth confinement}
\label{sec:finite_well}

The two-band Hamiltonian of Eq.~\eqref{eq:H} with a constant $V_s$ isolates the intervalley coupling responsible 
for the strain-induced splitting. A realistic quantum well, however, also includes a confining potential that is never infinitely deep.
We examine this effect numerically and show that the finite height of the confining potential primarily results in a slight 
renormalization of the effective width $L$ of the quantum well. Thus, the resulting valley splitting closely resembles 
that of an infinite well, with $L$ replaced by its renormalized value.

The hard-wall boundary condition $\psi(\pm L/2)=0$ underlying
Eq.~\eqref{eq:splitting} is an idealization; a real quantum well has a
finite conduction-band offset $U_0$ at its boundaries, and the wave function
penetrates into the barriers over a finite depth. This penetration is not
easily incorporated into the variational ansatz of Sec.~\ref{sec:ansatz},
since the plane-wave eigenvalleyors $\chi_\pm$ that the ansatz is built from
are not solutions of the barrier region; we therefore treat this extension
numerically only, using the same particle-in-a-box basis and coupling-matrix
machinery as the hard-wall problem, augmented with a valley-independent step
potential outside the well region.

We replace the hard-wall well of length $L$ with a well of the same length
embedded in a wider simulation box of length $L_{\rm box}=1.6L$, with a
valley-independent step potential $U_0$ outside $|z-L_{\rm box}/2|<L/2$ and
hard walls only at the edges of the larger box, pushed far enough
away not to affect the two lowest states. The same particle-in-a-box basis
and coupling matrix used for the hard-wall problem
(Eq.~\eqref{eq:Hprime} and Appendix~\ref{app:derivation}) apply unchanged
on this larger domain; the only new ingredient is the matrix element of the
step potential $U_0(z)$ in this basis, computed by the same quadrature used
for the coupling matrix.

\begin{figure}[t]
    \centering
    \includegraphics[width=0.95\linewidth]{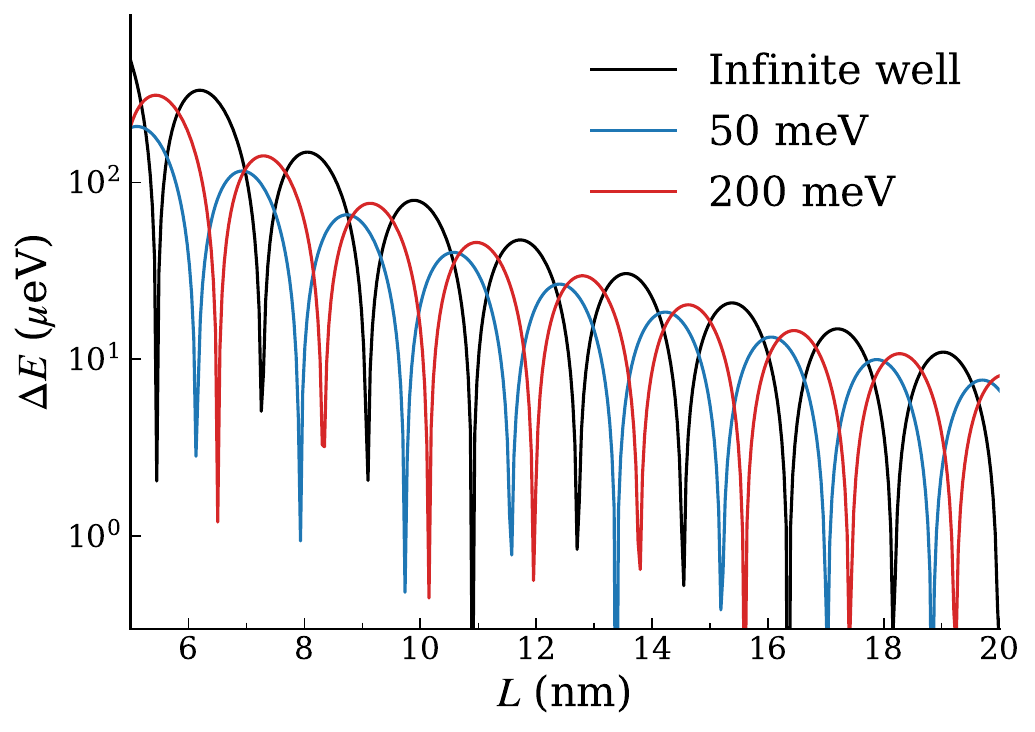}
    \caption{A finite barrier height preserves the oscillatory node
    structure of the hard-wall result but shifts it via wave function
    penetration into the barriers, while also systematically reducing the
    splitting magnitude at a given well width, especially for the
    shallower barrier: exact diagonalization for barrier heights
    $U_0=50$~meV (blue) and $U_0=200$~meV (red) compared to the
    hard-wall well of Sec.~\ref{sec:model} (solid black), at fixed
    $V_s=20$~meV over well widths $L=5$--$20$~nm for clarity.}
    \label{fig:finite_well}
\end{figure}

Figure~\ref{fig:finite_well} shows that finite barriers preserve the oscillatory structure but shift the nodes and peaks to larger effective widths while reducing the splitting, especially for smaller $U_0$.
This behavior follows from barrier penetration: the state experiences an effective width $L_{\rm eff}\approx L+2\ell_p(U_0)$, 
with $\ell_p$ decreasing as $U_0$ increases. Thus, to leading order Eq.~\eqref{eq:splitting} can be used to estimate 
$\Delta E(L_{\rm eff})$. Since $\Delta E(L)\propto\sin(k_{\rm min}L)/L^3$ at large $L$, the replacement $L\to L_{\rm eff}$ 
shifts the oscillation phase and reduces its magnitude. 
This suggests tuning the quantum-well depth $U_0$ to phase-shift the oscillations of the valley splitting, thereby 
allowing the system to be calibrated away from the nodes.

\section{Conclusion}
\label{sec:conclusion}
The valleyor basis, which corresponds to formulating the effective $k.p$ Hamiltonian around the $X$ point as in Eq.~\eqref{eq:H}, 
offers not only a significant computational advantage by eliminating unnecessary oscillatory factors, but also provides non-perturbative 
insight into the effects of shear strain and wiggle-well potentials across different regimes. The simplicity of this basis enables us 
to construct an accurate variational ansatz that captures the periodic dependence of the valley splitting on shear strain, as well as an 
``asymptotic freedom'' in the wiggle-well potential. In the strong-wiggle-well-amplitude limit, this latter behavior leads, counterintuitively, 
to valley splitting that is determined predominantly by the properties of the empty box, with the wiggle-well potential providing only a 
subleading correction.

 Our variational theory of valley splitting under uniform shear strain is based on plane-wave valleyors dressed by a hard-wall envelope.
The resulting closed-form splitting,
Eq.~\eqref{eq:splitting}, reproduces the numerical
$|\sin(k_{\rm min}L)|$ dependence and remains within $\sim1\%$ of exact
diagonalization for realistic $V_s\lesssim30$~meV. The remaining error arises
from the rigid single-harmonic envelope at much stronger coupling. 
The valley-density analysis further clarifies the structure of the resulting states, 
showing that the two valleys remain equally populated while coupled.  
The full two-component valleyor structure of the ansatz lets us verify that the two lowest-lying states possess well-defined valley polarization.
Finite barriers pose no difficulty as they just shift the node positions through wave function penetration and slightly reduce
the splitting without removing its oscillatory character.

The oscillatory dependence of the valley splitting on $\varphi=k_{\rm min}L$ suggests a strategy for optimizing 
shear strain for a fixed $L$ (or $L_{\rm eff}$ for finite wells): tune $\varepsilon_{xy}$ to bypass the canyon (nodes) associated with 
the nodes of $\sin\varphi$. In this way, the valley splitting can be enhanced by an order of magnitude compared to the 
left side of the first  node, as illustrated in Fig.~\ref{fig:splitting_vs_V1}. 
Our variational ansatz clarifies that the natural energy scale for the strain energy $V_s$ is the $X$-point energy $E_1$ 
measured relative to the parabolic minima. It also provides a simple rule of thumb for calibrating the shear strain away 
from the nodes and toward its optimal values. Specifically, one can start from the smallest possible values of $v_s^{\rm opt}$ 
given by Eq.~\eqref{eq:vs_opt}, which correspond to the integer $n$ in Eq.~\eqref{eq:nopt}.

For the WW coupling $V_w\cos(k_wz)\tau_1$, exact diagonalization shows a broad
resonance at $k_w=2k_1$ with diffraction-like sidelobes shown in Fig.~\ref{fig:splitting_vs_kw}. 
In the resonant frame, the low-energy physics crosses from nonorthogonal valley states away from
resonance to orbital quantization near resonance; in the strong-coupling
limit where the WW potential strength $V_w$ is larger than the box energy separation $E_\kappa$,
the valley splitting is dominated by empty box energy levels $E_\kappa$ given in Eq.~\eqref{eq:deltaE_max} with a 
subleading correction proportional to $q^2/V_w$ as in Eq.~\eqref{eq:wiggle_strongV}. 
These results, enabled by the valleyor representation, provide an analytical framework for interpreting the numerical wiggle-well spectrum.

\begin{acknowledgments}
S.A.J. thanks David P. DiVincenzo for fruitful discussions.  S.A.J was supported by EIN Quantum NRW. 
\end{acknowledgments}

\begin{appendices}
\section{Derivation of the variational energy}
\label{app:derivation}

This appendix gives the full evaluation of
$E_\pm=\bra{\psi_\pm}H\ket{\psi_\pm}$ and of the normalization
$|\mathcal{N}_\pm|^2$ summarized in Sec.~\ref{sec:ansatz}.

Define $f_\pm(z) = \cos(\kappa z)e^{\pm ik_{\rm min}z}$, so that
$\psi_\pm(z) = (\mathcal{N}_\pm/\sqrt L)\left(f_+(z)\chi_+ \pm f_-(z)\chi_-\right)$.
A direct calculation gives
\begin{align}
    & p_z f_\pm(z)= \hbar\left(i\kappa\sin(\kappa z)\pm k_{\rm min}\cos(\kappa z)\right)e^{\pm ik_{\rm min}z},\\
    & p_z^2 f_\pm(z) =\\ 
    &\hbar^2\left((\kappa^2+k_{\rm min}^2)\cos(\kappa z)\pm 2i\kappa k_{\rm min}\sin(\kappa z)\right)e^{\pm ik_{\rm min}z}\nonumber.
\end{align}
Acting with $H$ on $f_\pm\chi_\pm$ and using that $\chi_\pm$ is an
eigenvector of $\pm\frac{\hbar^2k_1k_{\rm min}}{m_\ell}\tau_3+V_s\tau_2$ with
eigenvalue $-\hbar^2k_1^2/m_\ell$ (from Eq.~\eqref{eq:kmin}), the terms
proportional to $\cos(\kappa z)e^{\pm ik_{\rm min}z}\chi_\pm$ combine into a constant
times $f_\pm\chi_\pm$, leaving
\begin{align}
    H(f_\pm\chi_\pm) =& \bar E_0 f_\pm\chi_\pm \\
    +&
    i\frac{\hbar^2k_1\kappa}{m_\ell}\sin(\kappa z)\left(\tau_3\pm\frac{k_{\rm min}}{k_1}\right)
    e^{\pm ik_{\rm min}z}\chi_\pm \notag,
\end{align}
with $\bar E_0 = \frac{\hbar^2}{2m_\ell}(\kappa^2+k_{\rm min}^2) - \frac{\hbar^2k_1^2}{m_\ell}$
(not to be confused with the constant $E_1$ of Eq.~\eqref{eq:Hprime}, which
plays no role here).
The residual term is the only piece that mixes $\chi_+$ and $\chi_-$ upon
taking the inner product with $\psi_\pm$. Using the identities
\begin{equation}
    \chi_+^\dagger\tau_3\chi_-=\chi_-^\dagger\tau_3\chi_+=0,\qquad
    \chi_\pm^\dagger\tau_3\chi_\pm=\mp\frac{k_{\rm min}}{k_1},
\end{equation}
which follow directly from Eq.~\eqref{eq:chi} (the phase $-i$ on the second component of $\chi_\pm$ cancels in every bilinear entering these identities), together with Eq.~\eqref{eq:chi_overlap}, the cross terms in
$\bra{\psi_\pm}H\ket{\psi_\pm}$ reduce to a single integral,
\begin{equation}
    E_\pm = \bar E_0 \pm \frac{\kappa V_sk_{\rm min}}{Lk_1^2}|\mathcal{N}_\pm|^2
    \int_{-L/2}^{L/2}\!dz\,\sin(2\kappa z)\sin(2k_{\rm min}z).
\end{equation}
Writing the product of sines as a sum of cosines and integrating over the
well gives
\begin{equation}
    E_\pm = \bar E_0 \pm \frac{\kappa^2V_sk_{\rm min}}{Lk_1^2(\kappa^2-k_{\rm min}^2)}
    |\mathcal{N}_\pm|^2 \sin(k_{\rm min}L),
\end{equation}
from which Eq.~\eqref{eq:splitting} follows immediately.

The normalization is fixed analogously, by integrating
Eq.~\eqref{eq:density} over the well:
\begin{align}
    1 &= \frac{2}{L}|\mathcal{N}_\pm|^2\!\int_{-L/2}^{L/2}\!dz\,\cos^2(\kappa z)
    \left(1\pm\frac{m_\ell V_s}{\hbar^2k_1^2}\cos(2k_{\rm min}z)\right)\nonumber\\
    &= |\mathcal{N}_\pm|^2\left(1\mp\frac{2}{L}
    \frac{m_\ell V_s\kappa^2}{2\hbar^2k_1^2(k_{\rm min}^3-k_{\rm min}\kappa^2)}\sin(k_{\rm min}L)\right),
\end{align}
which is Eq.~\eqref{eq:norm}.

\end{appendices}

\bibliography{references}

\end{document}